\documentclass[aps,reprint,superscriptaddress,amsmath,amssymb,floatfix]{revtex4-2}
\usepackage{graphicx}
\usepackage{float}
\usepackage{subcaption}
\usepackage[T1]{fontenc}
\usepackage{comment}
\usepackage{physics}
\usepackage{orcidlink}
\makeatletter
\renewcommand\@makecaption[2]{%
  \par
  \vskip\abovecaptionskip
  \begingroup
   \small\rmfamily
    \begingroup
     \samepage
     \flushing
     \let\footnote\@footnotemark@gobble
     \@make@capt@title{#1}{#2}\par
    \endgroup
  \endgroup
  \vskip\belowcaptionskip
}
\makeatother

\begin{document}

\preprint{APS/123-QED}
\title{Strain-stiffening critical exponents of fiber networks under uniaxial deformation}

\author{Atharva Pandit\,\orcidlink{0009-0000-6986-9773}}
\affiliation{Institut für Physik, Universität Augsburg, 86159 Augsburg, Germany}
\author{Fred C. MacKintosh\,\orcidlink{0000-0002-2607-9541}}
\affiliation{Department of Chemical and Biomolecular Engineering, Rice University, Houston, Texas 77005, USA}
\affiliation{Center for Theoretical Biological Physics, Rice University, Houston, Texas 77005, USA}
\affiliation{Department of Chemistry, Rice University, Houston, Texas 77005, USA}
\affiliation{Department of Physics and Astronomy, Rice University, Houston, Texas 77005, USA}
\author{Abhinav Sharma\,\orcidlink{0000-0002-6436-3826}}
\email[Contact author: ]{abhinav.sharma@uni-a.de}
\affiliation{Institut für Physik, Universität Augsburg, 86159 Augsburg, Germany}
\affiliation{Bereich Theorie der Polymere, Leibniz-Institut für Polymerforschung, 01069 Dresden, Germany}
\date{\today}

\begin{abstract}
Disordered fiber networks exhibit a floppy to rigid mechanical phase transition as a function of connectivity. Sub-isostatically connected networks can undergo this transition via straining. Critical exponents governing this transition have been estimated theoretically and by numerical simulations of various types of networks. In this study, we present improved results, achieved through a combination of refined numerical simulations, larger system sizes and incorporation of theoretical predictions for better post-simulation analysis. We also report the evolution of the critical strain and critical exponents as the network is sheared while being subjected to non-volume-preserving uniaxial deformations.
\end{abstract}

\maketitle
\section{Introduction}
\label{sec:intro}
Filamentous networks are ubiquitous in biology. At the intracellular level, actin filaments form disordered elastic networks that play a central role in controlling cell mechanics and motility~\cite{cytoskel_review}. At larger length scales, extracellular collagen networks constitute the primary structural components of tissues, providing mechanical integrity and structural support~\cite{collagen_review}. A hallmark property of such biopolymer networks is strain-stiffening: their elastic modulus increases dramatically under applied deformation~\cite{janmey1991, biogels2005, rheology_soft, chaudhuri2007, bausch2007, picu2011}. This pronounced nonlinear response suppresses large-scale deformations and is widely believed to protect cells and tissues against mechanical damage.

To understand the origin of this behavior, numerous studies have employed computational models of athermal, disordered elastic networks composed of semiflexible filaments subjected to external strain~\cite{frey2003,macK2003-1,macK2003-2,DasEMT,broedersz_review,lerner_sim}. These simulations quantify the mechanical response through stress--strain relations and differential moduli, consistently revealing strong nonlinear stiffening. Importantly, they also report highly heterogeneous and collective deformation fields, indicating that stiffening arises from network-scale cooperative rearrangements rather than from single-filament mechanics alone~\cite{SAlexander,frey2007,cross-link}.

\begin{figure}
    \centering
    \includegraphics[width=0.9\linewidth]{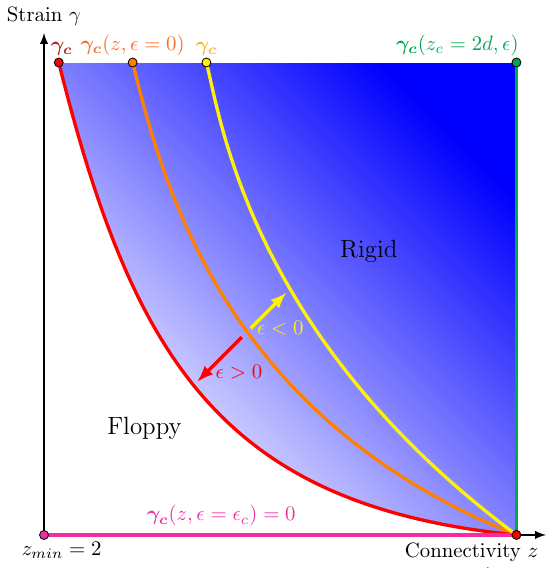}
    \caption{Sub-isostatic networks undergo a floppy to rigid transition at a critical strain $\gamma_c(z, \epsilon)$. Orange line shows the pure shear $\gamma_c(z)$ at which strain-stiffening occurs. Upon application of a uniaxial deformation $\epsilon$, the critical strain line shifts (red for extension and yellow for compression). Extending the network to $\epsilon=\epsilon_c$, the network is rigid even at zero shear (pink). For $z \le 2$, the network is not connected and can never become rigid. For $z \ge 2d$ (green), networks are always rigid.}
    \label{fig:illustration}
\end{figure}

The mechanical response of such disordered networks is fundamentally controlled by their connectivity. According to Maxwell counting, mechanical stability requires that the number of constraints balances the number of degrees of freedom, defining the isostatic threshold $z_c = 2d$ in $d$ dimensions~\cite{maxwell}. Networks with connectivity above this threshold are rigid even with central-force interactions alone, whereas subisostatic networks ($z < z_c$) are floppy in the absence of additional stabilizing interactions. Most biopolymer networks fall into this subisostatic regime, possessing fewer constraints than required for mechanical stability.

A key conceptual advance of the past decade has been the realization that such subisostatic networks can nevertheless become rigid under external deformation. In central-force networks, applied strain drives a transition from a floppy to a rigid state at a critical strain $\gamma_c$ that depends on connectivity ~\cite{elastic_regimes,stress_control,scalinglaw_abhi}. Additional interactions such as bending rigidity, thermal fluctuations, or active stresses stabilize the network below this threshold~\cite{mahadevan2008, broedersz2011, thermalfluct, activestress, shokef_safran}. For example, in networks with bending stiffness $\kappa$, the differential modulus scales as $K \sim \kappa$ for $\gamma < \gamma_c$, followed by a continuous transition to a stretching-dominated rigid phase above $\gamma_c$~\cite{exponent_abhi}.

Beyond static mechanical signatures, criticality in strained fiber networks has also been probed dynamically, where it manifests as critical slowing down, i.e., a strong divergence of stress-relaxation times as the system approaches the critical strain~\cite{slowdown}. The identification of strain-driven rigidity as a critical phenomenon has provided a unifying framework for understanding nonlinear stiffening in disordered networks. 
However, its universality class remains under active debate~\cite{threedim}. In particular, it is unclear whether the associated scaling behavior is governed by mean-field physics or reflects genuinely non-mean-field collective fluctuations~\cite{meanfield}.

Critical exponents $f$ and $\lambda$, with $\phi = \lambda + f$, describe the critical behavior, as $K$ exhibits power law scaling near $\gamma_c$, with $K \sim \kappa (\gamma_c -\gamma)^{-\lambda}$ in the subcritical regime, and $K \sim \mu (\gamma -\gamma_c)^f$ in the supercritical regime. Analytical approaches based on mean-field theory predict scaling relations between the critical exponents governing this transition, most notably the relation $\lambda = \phi - f$, and argue that these exponents are independent of the nature of deformation~\cite{scalinglaw_lerner, meanfield, effmedth}. Remarkably, numerical studies of networks subjected to shear, bulk expansion, or uniaxial strain have reported values of $\lambda$ that closely match this prediction, which has been interpreted as evidence that strain-induced rigidity belongs to a mean-field universality class. At the same time, simulations consistently find that the individual exponents $f$ and $\phi$ deviate from their mean-field values~\cite{bulkMod, uniaxial, compress_inclusion, f-phi, finite_size}, leaving unresolved whether the transition is truly mean-field.

In this work, we address this apparent paradox using large-scale simulations of subisostatic networks subjected to combined volumetric and shear deformations. We show that the exponent relation $\lambda = \phi - f$ remains robust across all conditions, with $\lambda$ always assuming the mean-field value, and $f$ varying systematically with connectivity and applied deformation. In particular, when networks are pre-compressed or expanded prior to shear, the critical behavior remains consistent with a strain-driven phase transition while the exponents evolve continuously. Our results demonstrate that agreement with the mean-field exponent relation does not imply mean-field criticality, and instead reveal a richer, deformation-dependent universality in strain-induced rigidity transitions.

\section{Strain-Driven Criticality and Scaling Theory}
\label{sec:theory}
We first briefly review the theoretical framework describing strain-driven criticality in subisostatic networks, including both general scaling theory and mean-field approach that have been proposed to describe the associated critical exponents. This provides the basis for the analysis and assumptions adopted in the present work.

Using real-space renormalisation, a scaling theory for networks undergoing strain-driven transitions has been developed~\cite{scalinglaw_abhi}. The critical behavior is captured through the exponents $f$, $\lambda$, and $\phi = f + \lambda$ via a Widom-like scaling form~\cite{widom}
\begin{equation}
K \approx \mu |\gamma - \gamma_c|^f \mathcal{G}_{\pm}(\tilde\kappa / |\gamma - \gamma_c|^\phi),
\label{eq:widom}
\end{equation}
where $\mu$ and $\kappa$ denote the stretching and bending moduli respectively, and $\pm$ corresponds to strain regimes above and below the critical strain $\gamma_c$. To characterise fiber elasticity, with $l_c$ as the average lattice spacing, $\tilde{\kappa} = \kappa / \mu l_c$ is defined as a reduced dimensionless fiber rigidity modulus, and is varied to stabilize networks in the limit $\tilde{\kappa} \to 0^+$. The scaling functions satisfy
\begin{subequations}\label{eq:Gdef}
\begin{gather}
\mathcal{G}_-(s \to 0) = \tilde\kappa / |\gamma - \gamma_c|^\phi, \\
\mathcal{G}_+(s \to 0) = 1, \\
\mathcal{G}_\pm(s \to \infty) = \tilde\kappa^{f/\phi}.
\end{gather}
\end{subequations}

The elastic energy per unit cell is denoted by $h(t,\kappa)$, where $t = \gamma - \gamma_c$ measures the distance from criticality. The critical point is approached as $t, \kappa \to 0$. Under a renormalisation step with scaling factor $L$ in $d$ dimensions, the energy transforms as
\begin{equation}
h(t,\kappa) = L^{-d} h(tL^x, \kappa L^y),
\label{eq:renorm_step}
\end{equation}
assuming parameter flows $t \to tL^x$ and $\kappa \to \kappa L^y$ with positive exponents $x,y$.

Denoting partial derivatives of $h$ with respect to $t$ and $\kappa$ as $h_{n,m}$, the stress $\sigma$ and differential stiffness $K$ follow from
\begin{subequations}\label{eq:stress_K_scale}
\begin{gather}
\sigma \sim \partial h / \partial t \sim L^{-d+x} h_{1,0}(tL^x, \kappa L^y), \\
K \sim \partial^2 h / \partial t^2 \sim L^{-d+2x} h_{2,0}(tL^x, \kappa L^y).
\end{gather}
\end{subequations}

By analogy with thermal critical phenomena, the correlation length scales as $\xi \sim L \sim |t|^{-\nu}$, yielding $\nu = 1/x$. This leads to the hyperscaling relation $f = d\nu - 2$. Above the critical strain, $h_{2,0}(1,s)$ approaches a constant for small $s$, giving $K \sim \mu |t|^f$, whereas below the transition $h_{2,0}(-1,s) \sim s$ leads to $K \sim \kappa |t|^{-\lambda}$, with $\lambda = \phi - f$.

Networks respond to applied strain through strongly non-affine internal rearrangements, which become increasingly pronounced near the critical strain and are analogous to divergent fluctuations in conventional critical systems~\cite{exponent_abhi, f-phi}. For an incremental strain $\delta \gamma$, non-affine fluctuations are quantified by
\begin{equation}
\delta \Gamma \sim \langle|\delta \mathbf{u} - \delta \mathbf{u}^A|^2\rangle / \delta \gamma^2,
\end{equation}
where $\delta \mathbf{u} - \delta \mathbf{u}^A$ measures the deviation from purely affine displacement. Since the elastic energy is minimized with respect to node displacements, for small bending rigidity one obtains $h \sim \kappa \delta \gamma^2 \delta \Gamma$. Consequently, non-affine fluctuations diverge as $\delta \Gamma \sim h_{2,1}(t,\kappa) \sim |t|^{-\lambda}$, governed by the same susceptibility-like exponent $\lambda$ that controls below-critical stiffness scaling.

Lerner \textit{et al.} have proposed an analytical approach with simplified description based on mean-field assumptions~\cite{scalinglaw_lerner, lerner_sim}. The mean-field theory focusses primarily on the below-critical regime, where the stiffness follows scaling relations such as
\begin{subequations}
\label{eq:lerner_scaling}
\begin{gather}
K \sim \mathcal{F}((\gamma_c - \gamma)/\delta\gamma_*(\kappa)), \\
K \sim \kappa (\gamma_c - \gamma)^{-3/2}.
\end{gather}
\end{subequations}
The latter expression corresponds directly to the below-critical branch of the Widom scaling form when the exponent relation $f - \phi = -\lambda = -3/2$ is imposed.

Analytical predictions and numerical studies across diverse network models have consistently reported values of $\lambda$ close to $3/2$, suggesting that this exponent relation may be robust and largely independent of microscopic details. Motivated by this apparent universality, in the present work we adopt $\lambda = 3/2$ as an empirical constraint and treat it as fixed across all systems and deformation protocols. The scaling relation $\phi = f + \lambda$ then reduces the critical behavior to a single independent exponent $f$, which we determine from large-scale simulations. This strategy allows us to directly assess whether strain-driven rigidity transitions exhibit genuinely non-mean-field critical behavior despite the apparent mean-field value of the exponent $\lambda$.

\section{Network Simulation and Mechanical Analysis}
\label{sec:methods}
Typically, disordered fiber networks are modelled using athermal lattice-based as well as off-lattice based models. It has been shown that both modelling approaches are equivalent, as long as the networks are sub-isostatic and have central force interactions~\cite{elastic_regimes}. In this study, we only work with 2D triangular lattice networks due to their overall simplicity, scalability and computational efficiency. The network generation protocol is described below. We assume no loss of generality with respect to other network architectures.

\begin{figure}[t]
    \centering
    \begin{subfigure}{0.333\columnwidth}
    {
        \includegraphics[width=\linewidth]{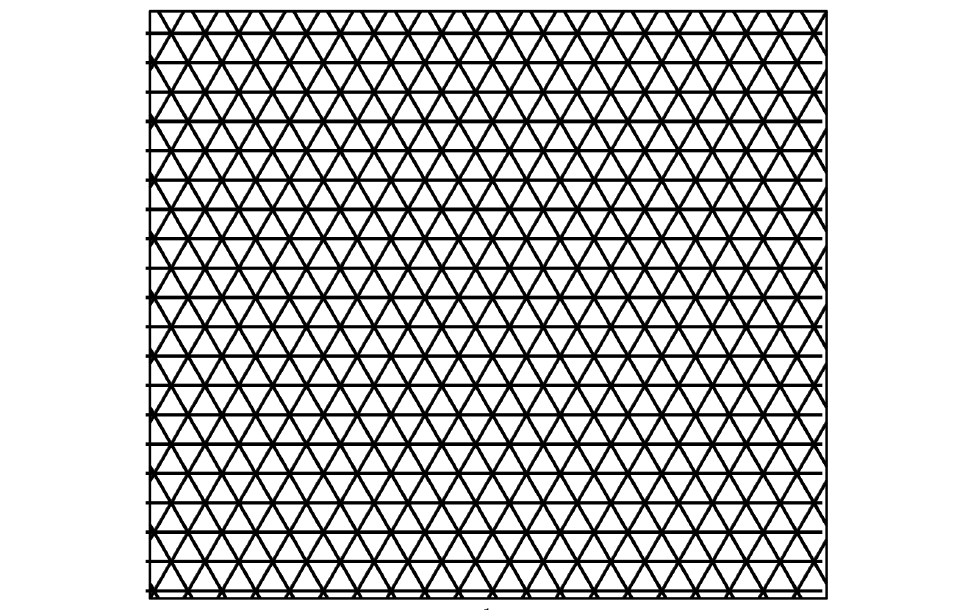}
        \caption{}
        \label{fig:Net4.0}
    }
    \end{subfigure}%
    \begin{subfigure}{0.333\columnwidth}
    {
        \includegraphics[width=\linewidth]{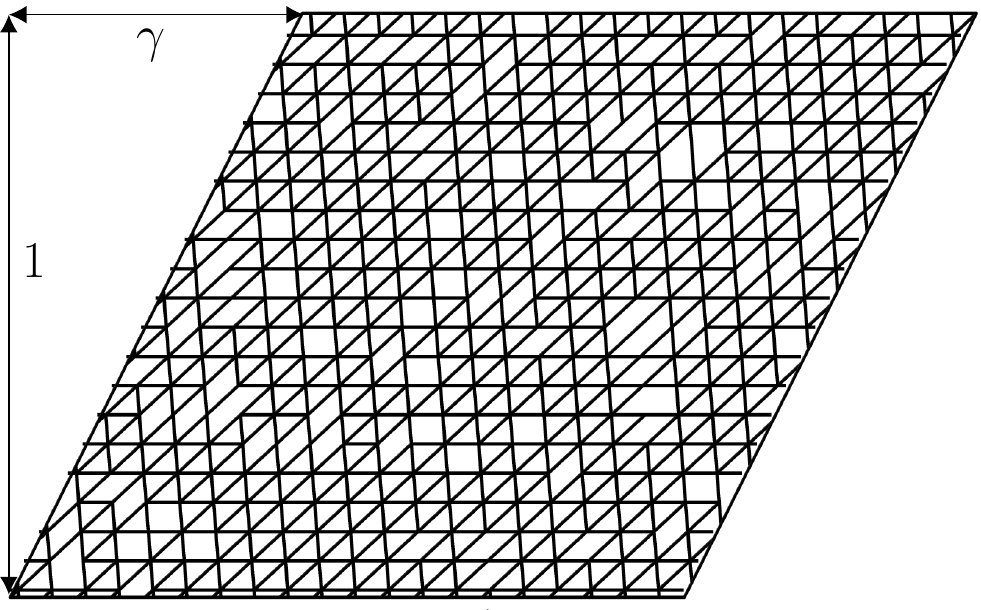}
        \caption{}
        \label{fig:Net3.6}
    }
    \end{subfigure}%
    \begin{subfigure}{0.333\columnwidth}
    {
        \includegraphics[width=\linewidth]{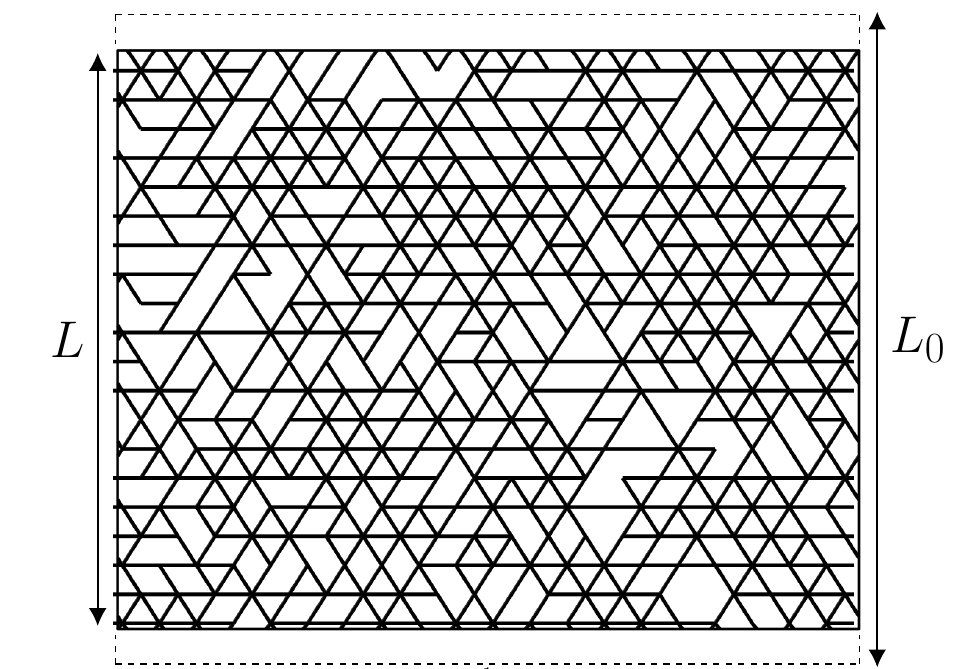}
        \caption{}
        \label{fig:Net3.2}
    }
    \end{subfigure}
    \caption{Phantomised 2D triangular lattices diluted to sub-isostatic connectivities and strained. (a) $z_c=4.0$ at zero strain. (b) $z=3.6$ under pure shear $\gamma$. (c) $z=3.2$ under uniaxial deformation $\epsilon= (L-L_0)/L_0$.}
    \label{fig:networks}
\end{figure}

Fig.~\ref{fig:networks} shows a 2D simulation box of $W \times W$ nodes arranged in a triangular lattice with nodes connected by bonds, with Lees-Edwards periodic boundary conditions imposed~\cite{LeesEdward}. Collinear bonds are considered to be part of a single filament. The networks are \textit{phantomised} by creating a binary cross-link with two randomly chosen filaments and keeping the third filament unhinged, resulting in a dilution from $z=6$ to $z=4$. To prevent system spanning filaments, one bond on each filament is cut. The networks are further diluted to the desired average connectivity by cutting random bonds. Nodes with zero or only one bond are removed as they do not contribute to system rigidity. Finally, a node is inserted at the centre of each bond to represent \textit{buckling} or the first bending mode.

Bonds are considered as Hookean springs with stretching modulus $\mu$ and bending interactions are considered along each filament between nearest neighbor bonds with bending modulus $\kappa$. The total network energy is given by
\begin{equation}
    H = \dfrac{\mu}{2}\sum\limits_{<ij>}\dfrac{(l_{ij}-l_{ij}^0)^2}{l_{ij}^0} + \dfrac{\kappa}{2}\sum\limits_{<ijk>}\dfrac{(\theta_{ijk}-\theta_{ijk}^0)^2}{l_{ijk}^0},
    \label{eq:energy}
\end{equation}
where the first term represents the stretching contribution due to springs between connected nodes $i$ and $j$, and the second term represents the bending contribution due to the angles between collinear bonds $ij$ and $jk$.

Once the network is set up, it can be driven to criticality by applying a pure shear strain (Fig.~\ref{fig:Net3.6}). In this study, we first apply a non-volume preserving uniaxial deformation $(\epsilon)$ to each network incrementally and quasi-statically (Fig.~\ref{fig:Net3.2}), while minimising it's energy at each step. $\epsilon > 0$ corresponds to an extension, while $\epsilon < 0$ is a compression. Once the network reaches the desired deformation, we shear it in both directions $(\gamma > 0$ and $\gamma < 0)$.

Energy minimisation at each step is done via employing the improved FIRE algorithm~\cite{fire, fire2}. We begin with a timestep $dt$ and iteratively integrate the particle positions until eventually we move uphill in the energy landscape, in which case we take a step back, reduce the timestep, reset the FIRE parameters according to the algorithm and continue minimising. The routine stops once $dt$ has been reduced by a factor of $100$ and no new energy minima were found in the five previous resets. Although it cannot be guaranteed if one has indeed reached the true energy minimum, the above mentioned procedure has proven to be robust in our tests.

We define network rigidity in terms of the differential shear modulus $K$, which is derived from the energy density of the network. For a network occupying volume $V$ (area in 2D), the total elastic energy per unit volume is given by $U = H/V$, where $H$ (Eq.~\eqref{eq:energy}) is minimized at each strain. Stiffness of a network can then be measured in terms of the shear modulus $K$, using $U$ and $\gamma$ as
\begin{equation}
    K = \pdv[2]{U}{\gamma}.
\end{equation}

We compute the differential non-affinity $d\Gamma$ in a system as
\begin{equation}
    d\Gamma = \dfrac{1}{N l_c^2 \delta\gamma^2} \sum_i || \delta u_i^{\mathrm{NA}} ||^2,
\end{equation}
where $N$ is the number of nodes, $l_c$ is the average bond length and $\delta u_i^{\mathrm{NA}}$ is the non-affine component of the displacement of node $i$ due to the incremental strain $\delta \gamma$.

To reduce noise in our results, we produce multiple realizations of each system at the same connectivity and system size, shear them in both directions, and report the average values. Unless stated otherwise, figures in this paper are of system size $W=200$ ($\sim 1.5-2 * 10^5$ nodes) and averaged over at least $4$ realizations with bidirectional shear.

\begin{figure}[b]
    \centering
    \begin{subfigure}{0.5\columnwidth}
    {
        \includegraphics[width=\linewidth]{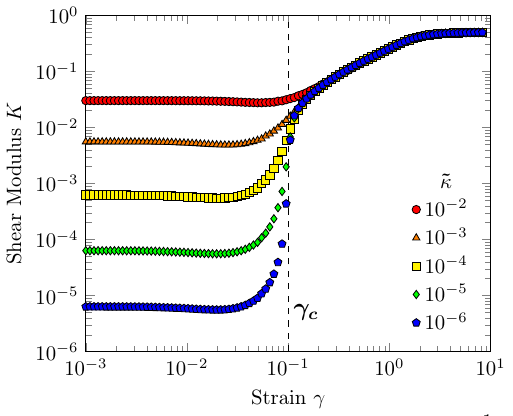}
        \caption{}
        \label{fig:SSA}
    }
    \end{subfigure}%
    \begin{subfigure}{0.5\columnwidth}
    {
        \includegraphics[width=\linewidth]{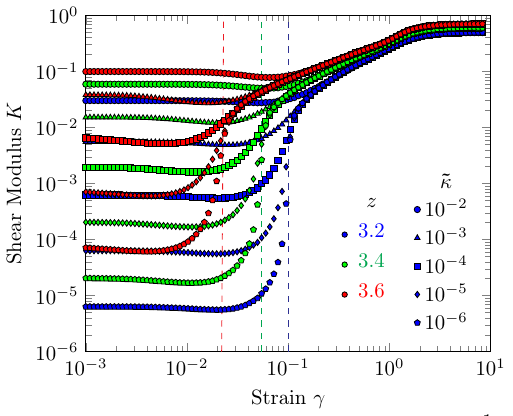}
        \caption{}
        \label{fig:SSB}
    }
    \end{subfigure}
    \begin{subfigure}{0.5\columnwidth}
    {
        \includegraphics[width=\linewidth]{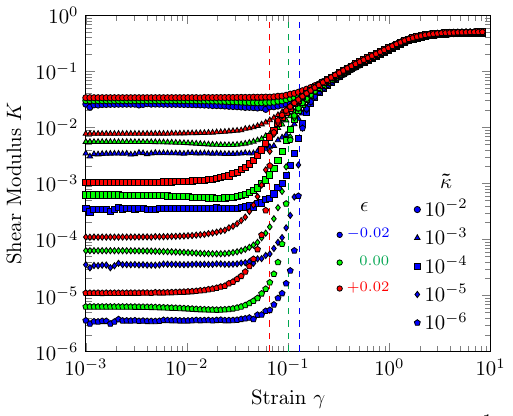}
        \caption{}
        \label{fig:SSC}
    }
    \end{subfigure}%
    \begin{subfigure}{0.5\columnwidth}
    {
        \includegraphics[width=\linewidth]{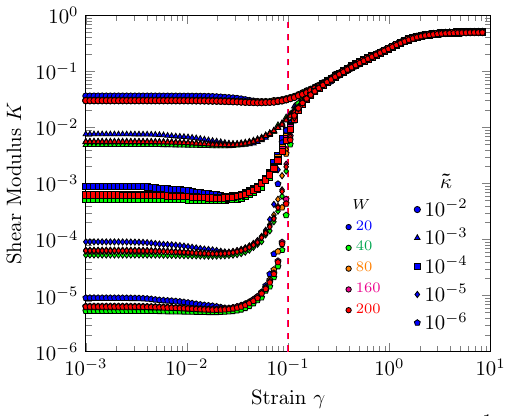}
        \caption{}
        \label{fig:SSD}
    }
    \end{subfigure}

    \caption{Differential shear modulus $K$ vs shear strain $\gamma$ with varying reduced bending modulus $\tilde\kappa$. (a) Dotted line shows the critical strain $\gamma_c$, where the bending dominated regime with $K \sim \tilde\kappa$ crosses over into the stretching dominated regime. (b) Earlier onset of rigidity as connectivity approaches $z_c=4$. (c) Networks under deformation show behavior similar to networks with different $z$. Compression $(\epsilon<0)$ delays criticality (resembling lower $z$) while extension $(\epsilon>0)$ advances it (resembling higher $z$). (d) Finite size effects do not shift the critical strain significantly. Initial softening of the modulus (buckling) is more prominent in smaller systems.}
    \label{fig:SS}
\end{figure}

\section{Mechanical Response and Critical Scaling}
\label{sec:results}

\subsection{Nonlinear Strain Stiffening}

\begin{figure}[b]
    \centering
    \begin{subfigure}{0.5\columnwidth}
    {
        \includegraphics[width=\linewidth]{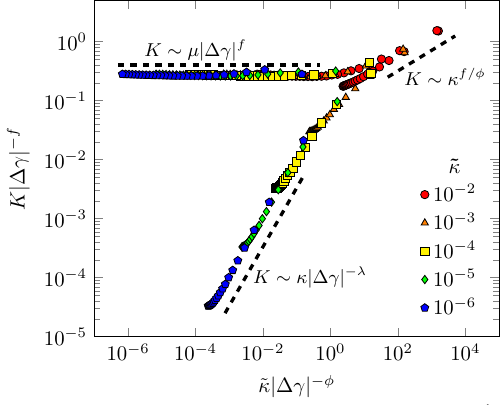}
        \caption{}
        \label{fig:colA}
    }
    \end{subfigure}%
    \begin{subfigure}{0.5\columnwidth}
    {
        \includegraphics[width=\linewidth]{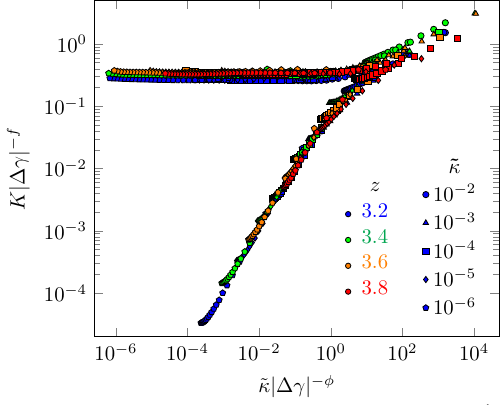}
        \caption{}
        \label{fig:colB}
    }
    \end{subfigure}
    \begin{subfigure}{0.5\columnwidth}
    {
        \includegraphics[width=\linewidth]{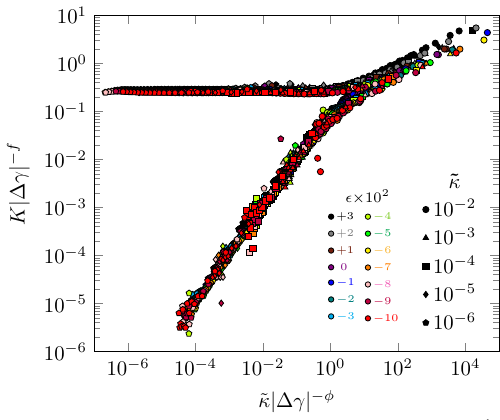}
        \caption{}
        \label{fig:colC}
    }
    \end{subfigure}%
    \begin{subfigure}{0.5\columnwidth}
    {
        \includegraphics[width=\linewidth]{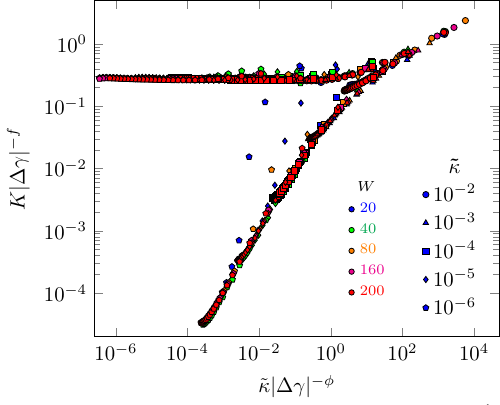}
        \caption{}
        \label{fig:colD}
    }
    \end{subfigure}

    \caption{Scaling collapse due to Widom-scaling (Eq.~\eqref{eq:widom}) with $\lambda=3/2$. (a) Three branches represent the different regimes of the rigidity transition (Eq.~\eqref{eq:Gdef}). (b) Widom-scaling across different connectivities at zero deformation. (c) Widom-scaling under compression and extension for $z = 3.2$. (d) Widom-scaling across finite size systems.}
    \label{fig:col}
\end{figure}

We begin by establishing that our simulations reproduce the well-known strain–stiffening behavior of subisostatic fiber networks reported in earlier studies. Fig.~\ref{fig:SS} summarises the shear modulus $K$ as a function of applied shear strain $\gamma$ under a wide range of conditions.  Fig.~\ref{fig:SSA} shows $K(\gamma)$ for a fixed network connectivity while varying the bending rigidity $\tilde\kappa$. Networks remain bending–dominated at small strain and exhibit a pronounced stiffening transition at a critical strain $\gamma_c$, beyond which stretching modes dominate. Increasing $\tilde\kappa$ shifts the onset of stiffening and raises the low–strain modulus, but the overall qualitative behavior remains unchanged.  
In Fig.~\ref{fig:SSB}, we plot the modulus–strain curves for different network connectivities at fixed bending rigidities. The results demonstrate that decreasing connectivity enhances the nonlinear response and shifts $\gamma_c$ to larger values. Fig.~\ref{fig:SSC} shows the effect of imposing an initial prestrain, either compressive or extensional, prior to applying shear. Pre-compression reduces the initial stiffness and increases the apparent critical strain, while pre-extension has the opposite effect.
Fig.~\ref{fig:SSD} presents $K(\gamma)$ for different system sizes at fixed $\tilde\kappa$. While the stiffening behavior is robust, smaller systems exhibit stronger fluctuations near $\gamma_c$, highlighting the growing importance of finite–size effects close to criticality. 

\subsection{Widom-Like Scaling of the Elastic Modulus}

Having established the basic stiffening phenomenology, we next examine the critical scaling of the shear modulus. The data in Fig.~\ref{fig:SS} can be collapsed using a Widom–like scaling form, Eq.~\eqref{eq:widom}, which captures the crossover between bending–dominated and stretching–dominated regimes near the critical strain. In performing this collapse, we fix the susceptibility–like exponent to the theoretically motivated value $\lambda = 3/2$, and treat the exponent $f$ as the only free parameter. This procedure significantly reduces the ambiguity associated with multi–parameter fitting and allows for a more reliable determination of the critical scaling behavior. The resulting collapses (Fig.~\ref{fig:col}), demonstrate excellent agreement across connectivities, prestrain conditions, bending rigidities, and system sizes, confirming the robustness of the Widom-type description. This analysis also clarifies a point of contention in the literature. Fig.~\ref{fig:col} shows that the Widom scaling consistently yields a positive $f$, in contrast to Ref.~\cite{scalinglaw_lerner} that states $f = 0$ is required to achieve consistency between Eqs.~\eqref{eq:widom} and~\eqref{eq:lerner_scaling}. The discrepancy arises because the theoretical framework of~\cite{scalinglaw_lerner, lerner_sim} is formulated only for the subcritical regime $(\gamma < \gamma_c)$, corresponding to the $\mathcal{G}_-$ branch of the scaling function, where $f$ does not appear and cannot be constrained.

It has also been argued in refs.~\cite{scalinglaw_lerner, lerner_sim} that the Widom-like scaling form is not valid. Their reasoning is based on two points. First, by imposing continuity of the scaling function at the critical strain, the Widom form predicts $K \sim \kappa^{f/\phi}$ at $\gamma_c$. However, they show that in strictly central-force networks ($\kappa = 0$) the modulus exhibits a discontinuous jump at the transition. This observation, however, does not contradict the scaling theory, because the case $\kappa = 0$ is fundamentally distinct from the limit $\kappa \to 0^+$. The Widom scaling applies to networks with finite bending rigidity, for which the modulus remains continuous and indeed follows the expected $K \sim \kappa^{f/\phi}$ scaling near criticality. Second, earlier tests of Widom scaling relied on large bending rigidities. This is not the case. In previous works \cite{scalinglaw_abhi, exponent_abhi}, as well the present work, we employ a range $\kappa = 10^{-2} - 10^{-6}$, which is comparable to the range $\kappa = 10^{-4} - 10^{-7}$ used in their own simulations. Within this regime we find that both the scaling functions $\mathcal{G}_\pm$ and the modulus vary continuously for all finite $\kappa$, and that the predicted scaling $K \sim \kappa^{f/\phi}$ is clearly observed in the immediate vicinity of $\gamma_c$ (see Figs.~\ref{fig:SSA} and~\ref{fig:colA}). 
Our results show that Widom-like scaling is consistent with numerical observations in the physically relevant limit of $\kappa \to 0^+$.

\subsection{Divergent Non-Affine Fluctuations}

\begin{figure}[b]
    \centering
    \begin{subfigure}{0.5\columnwidth}
        \includegraphics[width=\linewidth]{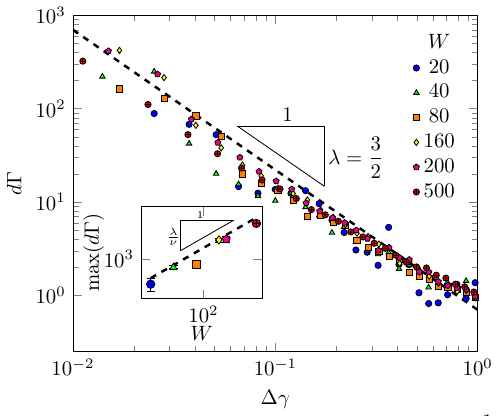}
        \caption{}
        \label{fig:NAfinite}
    \end{subfigure}%
    \begin{subfigure}{0.5\columnwidth}
        \includegraphics[width=\linewidth]{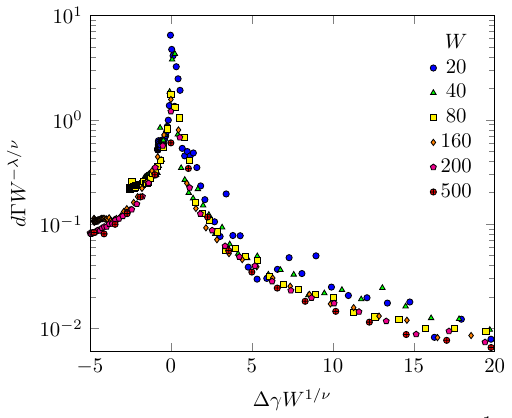}
        \caption{}
        \label{fig:FS}
    \end{subfigure}
    \begin{subfigure}{0.5\columnwidth}
        \includegraphics[width=\linewidth]{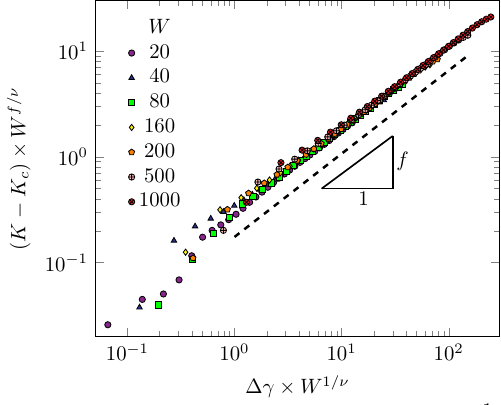}
        \caption{}
        \label{fig:FSWS}
    \end{subfigure}
    \caption{Near $\gamma_c$, non-affinity scales as $d\Gamma \sim |\Delta\gamma|^{-\lambda}$, in the limit $\kappa \to 0^+$. (a) $d\Gamma$ vs $\Delta \gamma = \gamma - \gamma_c$ for $\kappa=10^{-6}$ across systems sizes. Dotted line shows the slope $-3/2$. Inset: suppression of non-affine fluctuations due to finite size effects. Maximum of $\delta \Gamma$ scales as max$(\delta \Gamma) \sim W^{\lambda/\nu}$ with $\nu = 1.4 \pm 0.1$. (b) Finite-size scaling using $\lambda=3/2$ and $\nu = 1.4 \pm 0.1$, for networks with $z=3.2$ and $\kappa=10^{-6}$. (c) Finite-size scaling using $K-K_c$ for $\tilde\kappa=0$. For central force networks, the shear modulus $K$ exhibits a discontinous jump $K_c$ at $\gamma_c$.}
    \label{fig:NA}
\end{figure}

A central hallmark of a second-order phase transition is the divergence of a correlation length at criticality. In disordered fiber networks, this divergent length scale manifests through increasingly long-range, non-affine internal rearrangements as the critical strain is approached. These fluctuations can be quantified via the differential non-affinity, which measures the deviation of node displacements from purely affine deformation under an incremental strain. As discussed in Sec.~\ref{sec:theory}, the differential non-affinity is expected to diverge with the same susceptibility-like exponent $\lambda$ that governs the subcritical elastic response, $K \sim \kappa (\gamma_c - \gamma)^{-\lambda}$. Our results confirm this prediction. Fig.~\ref{fig:NAfinite} shows the differential non-affinity exhibits a clear power-law divergence consistent with $\lambda = 3/2$ across all systems, providing further justification for fixing this exponent in the scaling analysis.
At criticality, the correlation length becomes comparable to the system size, implying $W \sim \xi \sim |t|^{-\nu}$. Thus, the maximum value of $\delta \Gamma$ scales as $\delta \Gamma \sim W^{\lambda / \nu}$, shown in Fig.~\ref{fig:NAfinite} inset. Our determined value of $\nu = 1.4 \pm 0.1$ is also consistent with the hyperscaling relation in d-dim: $f = d \nu -2$.

Since the differential non-affinity scales as $d\Gamma \sim |t|^{-\lambda}$, this leads to the finite-size scaling relation $d\Gamma \sim W^{\lambda/\nu}$. The resulting scaling collapse, shown in Fig.~\ref{fig:FS}, demonstrates that the divergence of non-affine fluctuations is progressively suppressed by finite system size, as expected for a critical phenomenon.

Purely central force networks i.e.~$\tilde\kappa = 0$, exhibit a discontinuous jump $K_c$ in the shear modulus $K$ at $\gamma_c$. The stiffness near criticality scales as $K - K_c \sim W^{-f/\nu}$, leading to a scaling function of the form $K-K_c = W^{-f/\nu} \mathcal{F}(\Delta \gamma W^{1/\nu})$ \cite{finite_size}. $\mathcal{F}(s)$ is a constant for $s<1$ and $s^f$ for $s>1$, as validated in Fig.~\ref{fig:FSWS}.

Notably, the system sizes employed in our analysis, reaching $W=500$ ($\sim 9.3 \times10^5$ nodes for $z=3.2$) and $W=1000$ ($\sim 3.7 \times10^6$ nodes for $z=3.2$), substantially exceed those examined in any previous work, thereby strengthening the case for finite-size suppression effects, a key signature for second-order phase transitions. We note that Lerner \textit{et al.} have argued that their findings are inconsistent with the presence of such divergent differential non-affinity. Our results, however, clearly show both divergence and finite-size scaling of non-affine fluctuations, providing direct evidence for a growing correlation length and reinforcing the interpretation of strain-stiffening as a genuine second-order phase transition.

\subsection{Critical Exponents under Pre-Compression and Extension}

\begin{figure}[t]
    \centering
    \begin{subfigure}{0.9\columnwidth}
        \includegraphics[width=\linewidth]{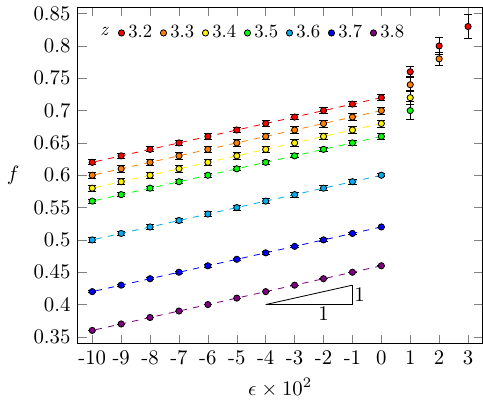}
        \caption{}
        \label{fig:compF}
    \end{subfigure}
    \begin{subfigure}{0.9\columnwidth}
        \includegraphics[width=\linewidth]{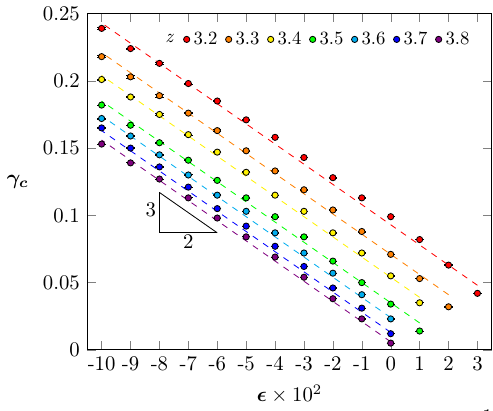}
        \caption{}
        \label{fig:compGAM}
    \end{subfigure}
    \caption{$f$ and $\gamma_c$ vs deformation $\epsilon$ averaged over system sizes $W=20, 40, 80, 160, 200$ with standard deviation as error bars. The variation is linear for $\epsilon < 0$ but shows non-linearity for $\epsilon>0$. System is rigid at zero shear for $\epsilon>\epsilon_c(z)$ and exhibits no criticality, hence $f$ and $\gamma_c$ are not reported.}
    \label{fig:comp}
\end{figure}

Fig.~\ref{fig:compF} summarizes the dependence of the critical exponents on connectivity and prestrain. At zero prestrain, we find that $f$ decreases monotonically with increasing connectivity $z$, while for a fixed connectivity $f$ decreases linearly under compression and increases nonlinearly under extension. These results show that although the critical exponent $\lambda$ remains fixed, the supercritical exponent $f$ is sensitive to both network structure and loading protocol. Because we impose the relation $\phi = f + 3/2$, the ratio $f/\phi$ varies only weakly across connectivities, consistent with the near overlap of the $K \sim \kappa^{f/\phi}$ branches in Figs.~\ref{fig:colB} and~\ref{fig:colC}. While it is often assumed that each network realization requires independent determination of $\gamma_c$, $f$, and $\phi$,~\cite{threedim,f-phi,finite_size} we find that the variability is in fact limited: the distribution of $\gamma_c$ is very narrow, particularly for larger systems. As shown in Fig.~\ref{fig:compGAM}, the averaged critical strain exhibits very small standard deviations across system sizes $W=20$--200 and follows a consistent linear dependence on prestrain $\epsilon$. Together with the absence of any systematic system-size dependence in the extracted exponents, these results demonstrate the robustness of the measured scaling behavior.

Previous numerical studies~\cite{threedim,f-phi,scalinglaw_abhi,finite_size} have reported a monotonic increase of $f$ and $\phi$ with connectivity, leading to a much larger variation in the ratio $f/\phi$. Our results however point to the opposite trend. While a detailed theoretical understanding of the exponent $f$ is still lacking, the consistency of our measurements across large system sizes and deformation protocols suggests that the observed behavior reflects intrinsic properties of the strain-driven criticality rather than finite-size or protocol-dependent effects.

\section{Discussion and Conclusion}
\label{sec:conclusion}
In this work, we have demonstrated that the Widom-like scaling framework provides a robust and unified description of strain-driven rigidity transitions in disordered fiber networks, even under general, non--volume-preserving deformations. By systematically studying shear combined with isotropic compression and extension, we show that networks subjected to mixed loading conditions continue to exhibit the same critical phenomenology that has previously been established only for pure shear or bulk deformations.

A key result of our study is that the susceptibility-like exponent $\lambda$ remains fixed at the value $\lambda = 3/2$ across all connectivities, deformation protocols, and system sizes considered. This finding is supported independently by scaling of the subcritical elastic response, divergence of differential non-affine fluctuations, and finite-size scaling collapse. Fixing $\lambda$ significantly reduces the number of free parameters in the Widom scaling analysis and enables reliable extraction of the exponent $f$ and the crossover exponent $\phi = f + \lambda$. Our results therefore clarify that the apparent agreement between mean-field predictions and numerical studies arises specifically from the robustness of $\lambda$, rather than from a genuinely mean-field nature of the transition.

In contrast, the exponent $f$, which governs the supercritical regime, exhibits systematic and nontrivial dependence on both network connectivity and deformation protocol. We find that $f$ decreases monotonically with increasing connectivity at zero prestrain, decreases linearly under compression, and increases nonlinearly under extension. These trends demonstrate that, unlike $\lambda$, the exponent $f$ is not universal and reflects the detailed mechanics of network rearrangements above the critical strain. At present, there is no theoretical framework capable of predicting $f$ from first principles.

We further show that the critical strain $\gamma_c$ is not strongly sample-dependent: its distribution is narrow and becomes increasingly well-defined with system size. Moreover, $\gamma_c$ shifts systematically under prestrain, following a linear trend with volumetric deformation. These observations suggest that strain-driven rigidity transitions may be governed by a multi-parameter critical surface in the space of connectivity, shear strain, and volumetric deformation.


Several open questions remain. In particular, the nonlinear regime under tensile prestrain ($0 < \epsilon < \epsilon_c$), where $f$ exhibits strong nonlinearity, requires further investigation. It will also be important to test the generality of the present results in three-dimensional networks and across a broader range of network architectures.

\begin{acknowledgments}
\label{sec:acknowlege}
A.S.~acknowledges support by Deutsche Forschungsgemeinschaft (DFG) within Project No.~SH 1275/5-1. A.P.~thanks F.~Faedi and B.~Valecha for helpful discussions.
\end{acknowledgments}

\section*{Data Availability}
The codes to reproduce the findings of this article are publicly available at~\cite{repo}.

\bibliography{ref.bib}
\end{document}